\documentclass[12pt,dvips]{article}

\usepackage{amsmath,amssymb,exscale}
\usepackage{array,multicol}
\usepackage{afterpage,float,flafter}
\usepackage{epsfig,rotating,pifont}
\usepackage{cite}
\@addtoreset{equation}{section} \makeatother

\def\beq{\begin{eqnarray}}
\def\eeq{\end{eqnarray}}

\def\lsim{\mathrel{\rlap{\lower3pt\hbox{\hskip0pt$\sim$}}
    \raise1pt\hbox{$<$}}}         
\def\gsim{\mathrel{\rlap{\lower4pt\hbox{\hskip1pt$\sim$}}
    \raise1pt\hbox{$>$}}}         

\title{
\vspace{1cm}
\Large\textbf{Predictive Power of  Strong Coupling in Theories with Large Distance Modified Gravity  }
\vspace*{.5cm}
\author{
\large \textbf{Gia Dvali\footnote{email: gd23@nyu.edu}}\\
\emph{Center for Cosmology and Particle Physics,}\\\emph{Department of Physics, New York University}\\
\emph{4 Washington Place, New York, NY 10003}}}

\date{}
\begin{document}
\maketitle \thispagestyle{empty} \vspace*{.5cm}

\begin{abstract}
 We consider theories that modify gravity  at cosmological distances, and show that any such theory  must exhibit a
 {\it strong coupling  phenomenon}, or else it is either inconsistent or is already ruled out 
 by the solar system observations.   We show that all the ghost-free theories that modify dynamics 
 of spin-2 graviton on asymptotically flat backgrounds,  automatically have this property. 
Due to the strong coupling effect, modification of the gravitational force  is source-dependent, and for lighter sources sets in at shorter distances.  This universal feature makes modified gravity theories predictive and  potentially testable not only by cosmological observations, but also by precision gravitational measurements at scales much shorter than the current cosmological horizon. 
We give a simple parametrization of consistent large distance  modified gravity  theories and 
their predicted deviations from the Einsteinian metric  near  the gravitating sources.

\end{abstract}

\newpage
\renewcommand{\thepage}{\arabic{page}}
\setcounter{page}{1}

\section{Introduction}

  The observed accelerated expansion of the Universe\cite{cc} could result \cite{ddg} from a  breakdown of the standard laws  of gravity at very large distances, in particular,  caused  by 
extra dimensions becoming gravitationally-accessible at distances of order the current cosmic horizon\cite{dgp}. 
  
     In this study, we shall formulate some very general properties 
  of large distance modified gravity  theories, that follow from the consistency requirements based on 4D effective field theory considerations.   These arguments allow to give a simple parametrization and display some very general {\it necessary} properties 
  of such theories, even without knowing their complete explicit form.  

 In this work, we shall be interested in theories that, {\it i)} are ghost free,  {\it ii)} admit the weak  field linearized  expansion on asymptotically flat backgrounds, and {\it iii)}  modify  Newtonian
 dynamics at a certain crossover  scale, $r_c$.  Motivated by cosmological considerations one is tempted 
 (and usually has to) to take $r_c$ of the order of the current cosmological horizon 
 $H_0^{-1} \sim 10^{28}$cm, but in most of our discussions we shall keep $r_c$ as a free parameter.  

 There exists an unique class of ghost-free  linearized theories of  a spin-2 field with the above properties. These theories have a tensorial structure  of the  Pauli-Fierz massive graviton, and are  described by the following equation  
\begin{equation}
\label{pf}
 \mathcal{E}^{\alpha\beta}_{\mu\nu} h_{\alpha\beta}\, 
\, - \, m^2(\square)\, (h_{\mu\nu} \, - \, \eta_{\mu\nu} h) \,  = \, T_{\mu\nu}. 
\end{equation}
 Here,
\begin{equation}
\label{einstein}
\mathcal{E}^{\alpha\beta}_{\mu\nu} h_{\alpha\beta}\, = \, \square h_{\mu\nu} \, - 
\, \square \eta_{\mu\nu} h  \, - \, \partial^{\alpha}\partial_{\mu} h_{\alpha\nu} \, -\, 
\partial^{\alpha} \partial_{\nu} h_{\alpha\mu} \, + \, 
\eta_{\mu\nu} \partial^{\alpha}\partial^{\beta}h_{\alpha\beta}\, + \, \partial_{\mu}\partial_{\nu} h
\end{equation}
is the linearized Einstein's tensor, and as usual $h \equiv  \eta_{\alpha\beta}h^{\alpha\beta}$.
$T_{\mu\nu}$ is a conserved source.  The gravitational coupling constant ($8\pi G_N$) is set equal to one, 
for simplicity.
 The difference of the above theory from the
Pauli-Fierz massive graviton is, that in our case $m^2(\square)$ is not a constant, but 
a more general operator, that depends on $\square$.  In order to modify Newtonian dynamics 
at scale $r_c$, $m^2(\square)$ must dominate over the first Einsteinian term 
at the scales $r \gg r_c$, or equivalently for the momenta $k \ll r_c^{-1}$. This implies that to the leading order 
\begin{equation}
\label{m}
m^2(\square) \simeq  r_c^{2(\alpha \, -\, 1)}\square^{\alpha\,},
\end{equation}
with $\alpha \, < \, 1$. 

The class of theories of interest thus can be parameterized by a single continuous parameter $\alpha$. 
As we shall see in a moment, unitarity further restricts $\alpha$  from below by demanding $\alpha \,\geqslant \, 0$. 
The  limiting case $\alpha\,  = \, 0$ corresponds to the massive Pauli-Fierz graviton.  

 Currently, the only  theory that has a known ghost-free generally-covariant non-linear completion 
 is $\alpha=1/2$, which is a five-dimensional brane world model (DGP-model) \cite{dgp}. 
 The limiting case $\alpha\, = \, 0$ for massive graviton,\footnote{Note,  $\alpha\, = \, 0$ 
 is not necessarily a massive gravity, since subleading  corrections to $m^2(\square)$ can further 
 distinguish the theories.  For instance, it is likely\cite{seesaw} that higher dimensional generalizations\cite{dg,ign}  of \cite{dgp}, should also correspond to $\alpha \, = \, 0$ case. }  
  unfortunately,  has no known consistent non-linear completion, and it is  very likely that such completions are impossible\cite{bd,gg}, at least in theories with finite number of states.
 
 The non-linear completions for theories with other values of $\alpha$ have not been studied.  It is not the purpose of the present study to discover such completions.  Instead, we shall 
 display some necessary conditions that such completions (if they exist) must satisfy. 
As we shall see, the effective  field theory considerations combined with consistency assumptions give us a powerful tool for predicting some of the general properties of the theories, even without 
knowing all the details of the underlying dynamics. 

 The central  role in our consideration is played by the {\it strong coupling} phenomenon
 of extra graviton polarizations, discovered
 in \cite{ddgv} both for massive gravity ($\alpha \, = \, 0$) and for \cite{dgp} ($\alpha\, = \, 1/2$). 
 We shall show, that this phenomenon must be exhibited by theories with  arbitrary values of $\alpha$ 
 from the unitarity range $ 0 \, \leqslant \, \alpha \, \leqslant\,  1$.  Due to this strong coupling, a gravitating 
 source of mass $M$, on top of the ordinary Schwarzschild radius  $r_g \, \equiv \, 2 G_N M$, 
 acquires  a second (intermediate) physical distance  $r_g \, \leqslant \,  r_* \, \leqslant \, r_c$, 
 which for extra graviton polarizations plays the role somewhat similar to the Schwarzschild  horizon. 
 The new scale $r_*$ depends both  on $r_g$ and $r_c$,  and thus, varies from source to source. 
 
 The above fact  predicts the following modification of the gravitational 
 potential for an arbitrary source localized within $r_*$-radius 
\begin{equation}
\label{centralcor}
\delta \, \sim \, \left ({r \over r_c} \right )^{2(1-\alpha)} \sqrt{r \over r_g}. 
\end{equation}
Because of  this  predictive power, the theories of modified gravity 
are very restrictive and potentially testable by precision gravitational measurements.  Our analysis
complements and solidifies the earlier study of \cite{dgz, nobel},  in which the corrections (\ref{centralcor}) where already suggested. 

Notice, that for the cosmologically interesting case $r_c \,  = \, H_0^{-1}$, the entire observable  Universe 
is the only source for which $r_g \, = \, H_0^{-1}\, = \, r_*$.  Therefore,  the concept of $r_*$ is crucial for understanding stability (absence of ghosts or tachions)  of  cosmological backgrounds, 
thus, invalidating many perturbative approaches, as the sources of size $r_*$ have to be studied 
nonperturbatively.

 \section{Unitarity Constraints}

 We shall first derive the unitarity lower bound on $\alpha$.  
For this,  consider a one-graviton exchange amplitude between the two conserved sources
($T_{\mu\nu}$ and $T_{\mu\nu}'$) 
 \begin{equation}
\label{amplitude}
({\rm Amplitude})_{h} \, \propto \, {T'_{\mu\nu}T^{\mu\nu} \, - {1 \over 3} T_{\mu}^{\mu} T_{\nu}^{'\nu} 
\over   \square\, - \, m^2(\square)}.
\end{equation}
The scalar part of the (Euclidean) graviton
propagator then has the following form 
\begin{equation}
\label{scalarprop}
\Delta(k^2) \, = \, {1
\over   k^2\, + \, m^2(k^2) }.
\end{equation}
As said above, modification of gravity at large distances implies that the denominator is dominated by the second term, for 
$k \rightarrow 0$.  To see how unitarity constrains such a behavior, let the spectral representation 
of the propagator $\Delta(k^2)$ be 
\begin{equation}
\label{spectral}
\Delta(k^2) \, = \, \int_0^{\infty} ds\,  { \rho (s) \over k^2 + s},
\end{equation}
where $\rho(s)$ is a bounded spectral function.  The absence of the negative norm states demands that 
$\rho(s)$ be a semi-positive definite function. Evaluating (\ref{spectral}) for the small $k$ and using parameterization 
(\ref{m}), we get
\begin{equation}
\label{smallk}
{1\over  r_c^{2(\alpha \, -\, 1)} k^{2\alpha\,}}\, = \, \int_0^{\infty} ds \,{ \rho (s) \over k^2 + s}.
\end{equation}
  Non-negativity of $\rho(s)$ implies that $\alpha$
cannot be negative.  In the opposite case, $\Delta(k^2)$ 
would be zero for $k^2=0$, which is impossible for non-negative $\rho(s)$.  Thus, from unitarity, 
it follows that $\alpha \, = \, 0$ is the lowest possible bound. 

 In general, branch cuts in  the propagator reflect the existence of the continuum of states, 
which could be interpreted as  the signature of extra dimension.  Thus, in a sense, large distance 
modification of gravity leads to the  existence of extra non-compact dimensions, as in \cite{dgp}.

\section{Extra Polarizations}

A graviton  satisfying the equation (\ref{pf}),  just as in the Palui-Fierz case,  propagates five degrees of freedom. These include,  two tensor, two vector and one scalar polarizations.  The extra polarizations (especially the scalar) play the central role in the strong coupling  phenomenon, and it is therefore useful to separate 
the `new' states from the `old'  (massless spin-2)  states. 
 
 We shall now  integrate the extra three polarizations out
and write down the effective equation for the two remaining  tensorial ones.  For this we shall first rewrite 
equation (\ref{pf}) in the manifestly gauge invariant form, using the St\"uckelberg method. 
We can rewrite $h_{\mu\nu}$ in the following form 
\begin{equation}
\label{vector}
h_{\mu\nu} \, = \, \hat{h}_{\mu\nu} \, + \, \partial_{\mu}A_{\nu} \, + \, \partial_{\nu} A_{\mu}, 
\end{equation}
where the St\"uckelberg field $A_{\mu}$ is the massive vector that carries two extra helicity-1 polarizations. The remaining extra scalar state resides partially in $A_{\mu}$ and partially in $\hat{h}_{\mu\nu}$. 
Written in terms of $\hat{h}_{\mu\nu}$ and $A_{\mu}$, 
  \begin{equation}
\label{pfs}
 \mathcal{E}^{\alpha\beta}_{\mu\nu} \hat{h}_{\alpha\beta}\, 
\, - \, m^2(\square) (\hat{h}_{\mu\nu} \, - \, \eta_{\mu\nu} \hat{h}  \, + \partial_{\mu}A_{\nu} \, + \, \partial_{\nu} A_{\mu} \, - \, 2 \eta_{\mu\nu} \partial^{\alpha}A_{\alpha}) \,  = \, T_{\mu\nu},
\end{equation}
the equation (\ref{pf}) becomes manifestly invariant under the following gauge transformation 
\begin{equation}
\label{gauge}
\hat{h}_{\mu\nu} \, \rightarrow \hat{h}_{\mu\nu}  \, +  \, \partial_{\mu}\xi_{\nu} \, + \, \partial_{\nu} \xi_{\mu},
~~~A_{\mu} \, \rightarrow \, A_{\mu} \,  - \, \xi_{\mu},
\end{equation}
where $\xi_{\mu}$ is the gauge parameter.
Note that the first, Einstein's term is unchanged under the replacement (\ref{vector}),  due to it's gauge invariance.  We now wish to integrate out $A_{\mu}$ through its equation of motion, 
  \begin{equation}
\label{Aequ}
\partial^{\mu} F_{\mu\nu} \, = \, - \, \partial^{\mu}  (\hat{h}_{\mu\nu} \, - \, \eta_{\mu\nu} \hat{h} ),
\end{equation}
where $F_{\mu\nu} \, \equiv \, \partial_{\mu}A_{\nu}  - \partial_{\nu}A_{\mu}$. 
Before solving for $A_{\mu}$, note that by taking a divergence from  the equation (\ref{Aequ})
we get the following constraint on $\hat{h}_{\mu\nu}$ 
 \begin{equation}
\label{hconstraint}
 \partial^{\mu} \partial^{\nu}  \hat{h}_{\mu\nu} \, -  \, \square \hat{h} \, = \, 0,
\end{equation}
which means that $\hat{h}_{\mu\nu}$ is representable in the form 
\begin{equation}
\label{tilderep}
\hat{h}_{\mu\nu} \, = \, \tilde{h}_{\mu\nu} \, + \, \eta_{\mu\nu} {1\over 3} \Pi_{\alpha\beta}\tilde{h}^{\alpha\beta},
\end{equation}
where $\Pi_{\alpha\beta}\, = \, {\partial_{\alpha}\partial_{\beta} \over \square} \, - \, \eta_{\alpha\beta}$ 
is the transverse projector.  $\tilde{h}_{\mu\nu}$ carries two degrees of freedom.
Notice that, since the last term in (\ref{tilderep}) is gauge invariant, under the gauge transformations 
(\ref{gauge}), $\tilde{h}_{\mu\nu}$ shifts in the same way as $\hat{h}_{\mu\nu}$ does.

Coming back to the equation (\ref{Aequ}) and solving it for $A_{\mu}$ we get 
 \begin{equation}
\label{asolution}
A_{\nu} \, = \, - \, {1 \over \square} \partial^{\mu}  (\hat{h}_{\mu\nu} \, - \, \eta_{\mu\nu} \hat{h} )
\, - \, \partial_{\nu} \Theta, 
\end{equation}
where $\Theta$ is arbitrary.  The gauge invariance demands that under the transformations 
(\ref{gauge}), $\Theta$ shifts in the following way
\begin{equation}
\label{thetagauge}
\Theta \, \rightarrow \, \Theta \, + \, {1 \over \square} \partial_{\alpha}\xi^{\alpha}.
\end{equation}
Substituting (\ref{asolution}) back to (\ref{pfs}),  expressing  
$\hat{h}_{\mu\nu}$ in terms of  $\tilde{h}_{\mu\nu}$ through  (\ref{tilderep}),
and choosing the gauge appropriately, 
 we can write the resulting effective equation for $\tilde{h}_{\mu\nu}$ in the following form
 \begin{equation}
\label{linear}
 \left ( 1 \, - \, {m^2(\square)\over \square} \right ) \mathcal{E}^{\alpha\beta}_{\mu\nu} \tilde{h}_{\alpha\beta}\, 
 = \, T_{\mu\nu},
\end{equation}
which has the tensorial structure identical to the linearized Einstein's equation. This latter fact  indicates that 
$\tilde{h}_{\mu\nu}$ indeed propagates only two tensor  polarizations, characteristic for the massless graviton.  

  Because $\tilde{h}_{\mu\nu}$ propagates only two degrees of freedom, the one-particle exchange amplitude between the two conserved sources $T_{\mu\nu},  T_{\mu\nu}'$ mediated 
by $\tilde{h}$ is
\begin{equation}
\label{amplitudepart}
({\rm Amplitude})_{\tilde{h}} \, \propto \, {T'_{\mu\nu}T^{\mu\nu} \, - {1 \over 2} T_{\mu}^{\mu} T_{\nu}^{'\nu} 
\over \square\,  - \, m^2(\square)},
\end{equation}
which continuously recovers the massless graviton result in the limit $m^2 \rightarrow 0$. 
This fact however {\it does not} avoid the well known van Dam-Veltman-Zakharov (vDVZ) discontinuity\cite{vDVZ}, because (\ref{amplitudepart})
is only a part of the full one-particle exchange amplitude.  This becomes immediately clear if we 
notice that the metric excitation that couples to the conserved $T_{\mu\nu}$ is $\hat{h}_{\mu\nu}$ (or equivalently  $h_{\mu\nu}$), which
depends on $\tilde{h}_{\mu\nu}$ through (\ref{tilderep}).  The full physical amplitude 
is generated by the latter combination of $\tilde{h}_{\mu\nu}$ and is equal to (\ref{amplitude}), 
which clearly exhibits the vDVZ-type discontinuity in the $m^2(\square)\, \rightarrow  \, 0$ limit.

\section{Strong Coupling of Extra States and Concept of $r_*$ }

We have seen that in the considered class of theories graviton should contain three extra 
`longitudinal' polarizations, which lead to vDVZ discontinuity in linearized theory. 
Hence, any theory of modified gravity that remains weakly coupled at the solar system distances, 
is automatically ruled out by the existing gravitational data. 

 Thus, the only point that could save such theories is the breakdown of the linearized 
approximation at the solar system distances, that is, the {\it strong coupling phenomenon} \cite{ddgv}. 
 In the other words, the only theories of modified gravity that can be  compatible 
with observations are the strongly coupled ones.  Any attempt of curing the strong coupling, 
will simply rule out the theory.  

 Fortunately, the same extra polarization that creates a worrisome discontinuity   in the 
 gravitational amplitude, also provides the strong coupling which invalidates the unwanted result
 at solar system scales. 
 We shall now generalize the results of \cite{ddgv}  to theories with  arbitrary $\alpha$.
 
Consider a theory of graviton that in the linearized approximation 
satisfies the equation (\ref{pf}).  We shall assume that this theory has a generally-covariant non-linear 
completion. The non-linear completion of the first term is the usual Einstein's tensor. The completion 
of the second  term is unknown for general $\alpha$. In fact, extrapolation from the only 
known case of $\alpha=1/2$\cite{dgp} tells us that  if a completion exists, it will probably require going beyond four-dimensions.  Discovering such a completion  is beyond our program. 
In fact, knowing the explicit form of this completion is unnecessary for our analysis, provided
it meets consistency  requirements listed above.   Then, we will be able to derive general 
properties of the strong coupling and the resulting predictions.

The propagator for the graviton satisfying equation (\ref{pf}) has the following  form,
\beq
\label{prop}
D_{\mu\nu ;\alpha\beta}\,=\, 
\left(
{1\over 2} \,\tilde\eta_{\mu\alpha} \tilde \eta_{\nu\beta}+
{1\over 2} \, \tilde\eta_{\mu\beta}  \tilde\eta_{\nu\alpha}-   
{1\over 3} \, \tilde\eta_{\mu\nu} \tilde \eta_{\alpha\beta}\right)\frac{1}{
\square\, - \, m^2(\square)}\,,
\label{5D}
\eeq  
where 
\begin{equation}
\label{etatilde}
\tilde\eta_{\mu\nu}\, \equiv \, \eta_{\mu\nu}\, - \,  {\partial_\mu \partial_\nu \over m^2(\square)}
\, = \, \eta_{\mu\nu}\, - \, r_c^{2(\alpha\, - \,1)} {\partial_\mu \partial_\nu  \over \square^{\alpha}}.
\end{equation}

 The existence of the terms that are singular in $r_c^{-1}$ is the most important fact. 
This singularity is precisely the source of the strong  coupling observed in \cite{ddgv} for
$\alpha\, =\, 0$ and $\alpha\, = \, 1/2$ cases.  The terms that are singular in $r_c^{-1}$ come from the additional,
scalar polarization state of the resonance  graviton. In the St\"uckelberg language given 
in (\ref{gauge}), this scalar  polarization resides partially in $A_{\mu}$ and partially
in the trace of $\hat{h}_{\mu\nu}$.  If we denote the canonically normalized  longitudinal  polarization by  $\chi$, then, ignoring the spin-1 helicity vector, the full metric fluctuation to the linear order
can be represented as 
 \begin{equation}
\label{handchi}
h_{\mu\nu} \, = \, \tilde{h}_{\mu\nu} \,  - \, {1\over 6} \eta_{\mu\nu}\chi \, + \,
 r_c^{2(1\,  - \alpha\,)} {\partial_\mu \partial_\nu  \over 3\square^{\alpha}}\chi.
\end{equation}
Notice that $\tilde{h}_{\mu\nu}$ is the same as in (\ref{tilderep}).
The same state is responsible for the 
extra attraction, that provides factor of $1/3$ in the one graviton exchange amplitude, 
as opposed to $1/2$ in the standard gravity, leading to vDVZ discontinuity.  

  The strong coupling of the longitudinal gravitons, however, has a profound effect on the discontinuity. 
The effect of the longitudinal gravitons  becomes suppressed
near the gravitating sources, where the linearized approximation  breaks down. 
Due to the strong coupling effects, the gravitating sources of mass $M$, on top of the usual 
Schwarzschild gravitational radius $r_g \equiv 2G_N \, M$, acquire the second physical radius, which we shall call $r_*$. Breakdown of the linearized approximation near gravitating sources, was first noticed 
in the context of $\alpha=0$ theory in \cite{av}, and the underlying strong coupling 
dynamics was uncovered in\cite{ddgv}.  The key point of the present discussion is that 
both the strong coupling, and the resulting $r_*$-scale is a property of theories with arbitrary 
$\alpha$. 

  For the longitudinal graviton $\chi$, the latter radius plays the role somewhat similar 
to  the one played by  $r_g$  
for the two usual tensor polarizations  $\tilde{h}_{\mu\nu}$.  Namely, due to the strong coupling,  at $r=r_*$ the non-linear self interactions of $\chi$ become  important, and the expansion in series of $G_N$ (that is, expansion in $r_g/r$) breaks down. 

 The concept of $r_*$ plays the central role in any large 
 distance modified theory of gravity, as we shall now discuss. 
 Consider a localized  static gravitating source $T_{\mu\nu} \, = \, \delta_{\mu}^0\delta_{\nu}^0 M
 \delta(r)$
 of gravitational radius $r_g$. 
 Then, sufficiently far  from the source, the linearized approximation  should be valid, and the metric 
 created by the source can be found in one graviton exchange approximation to linear order in 
 $G_N$,
\begin{equation}
\label{metricofM}
h_{\mu\nu}  \, = \, {\delta_{\mu}^0\delta_{\nu}^0\, - {1 \over 3} (\eta_{\mu\nu}\, - \, r_c^{2(1\, - \, \alpha\,)} {\partial_\mu \partial_\nu  \over \square^{\alpha}})
\over \square \,  - \, r_c^{2(\alpha\,  - \,1)} \square^{\alpha}}\,  r_g  \, 4\pi \delta(r).
\end{equation}
The term in the numerator,  that is singular in $1/r_c$,  vanishes when convoluted with any conserved test
source $T'_{\mu\nu}$, in accordance with (\ref{amplitude}). Hence,  at the distances below  $r_c$, 
the metric has an ${r_g \over r}$ form, but with a {\it wrong}  (scalar-tensor type)  tensorial structure, 
manifesting vDVZ discontinuity.    However, in nonlinear interactions, the singular in $1/r_c$ terms no longer vanish and in fact wash-out the linear effects, due to the strong coupling. The scale of the strong coupling  can be figured out by generalizing the analysis of \cite{ddgv, dgz}. The straightforward  power counting then 
gives that  the leading singularity in $r_c$ is of the order of $r_c^{4(1-\alpha)}$,
and comes from the trilinear  interaction of the longitudinal gravitons. 
For $\alpha\, = \, 0$ this reproduces the result of \cite{ddgv} (see also \cite{m2}).
For general $\alpha$ this vertex has a momentum  dependence of the form
\begin{equation}
\label{vertex}
      (r_ck)^{4(1-\alpha)}k^2. 
\end{equation}
Then, the  scale $r_*$ corresponds to a distance from the source for which the contribution from the above trilinear vertex, becomes as important as the linear one given by (\ref{metricofM}).  
 It is obvious that the corresponding distance is given by 
 \begin{equation}
\label{rstar}
r_* \, = \, (r_c^{4(1-\alpha)}r_g)^{{1 \over 1 + 4(1-\alpha)}}.
\end{equation} 

For distances $r \ll r_*$ the correction to the Einsteinian  metric coming from the longitudinal gravitons 
is suppressed by powers of $r_*$. The leading behavior can be fixed from the two requirements. 
First,  $\chi(r)$  should become  of order $r_g/r_*$ at $r=r_*$, in order to match the linear regime  (\ref{metricofM})  outside the $r_*$-sphere.  Secondly, the solution inside $r_*$ must be possible to approximate
by the analytic  series in $r_c^{2(\alpha-1)}$. These two requirements fix the leading behavior as
\begin{equation}
\label{correction}
\chi (r \ll r_*) \, \sim \,  {r_g \over r_*} \left ({r \over r_*}\right )^{{3 \over 2} - 2 \alpha}, 
\end{equation} 
which gives the relative correction to the gravitational potential given by (\ref{centralcor}).

For $r_c \, = \, H_0^{-1}$, the predicted anomalous perihelion precession of a planetary orbit is given by 
\begin{equation}
\label{pre}
{\rm Perihelion~advance~per~orbit} ~~\, \sim \, \left (r H_0 \right )^{2(1-\alpha)} \sqrt{r \over r_g}, 
\end{equation}
 where $r_g$ is the gravitational radius of the binding star (or any other source), and $r$ is the radius of the 
 planetary orbit around it.  This expression reproduces the result of \cite{dgz}.

As pointed out in the latter paper, for some interesting values of $\alpha$  these corrections are strong 
enough to be tested in precision gravitational measurement experiments. For $\alpha = 1/2$ theory this was  also pointed out in \cite{ls},  with the correction from the cosmological 
self-accelerated  background \cite{cd} 
taken into the account. 

 In fact, existing Lunar Ranging 
constraints already rule out theories with $\alpha$ significantly above $1/2$, and the new generation of the improved accuracy measurements\cite{lr}, will probably test  the $\alpha \simeq 1/2$ case.   

Notice that for $\alpha =1/2$, (\ref{centralcor}) matches the known explicit solutions derived both in
$1/r_c$-expansion\cite{andrei}, as well as exactly\cite{gi}.

\section{Conclusions}

Possibility of large-distance modification of gravity is a fundamental question, motivated by 
the Dark Energy Problem. 
  In this paper we tried to stress the important role played by the {\it strong coupling phenomenon}  in
  this class of theories.  In particular, it was already appreciated \cite{ddgv}, \cite{nobel} that the strong coupling is the only  cure that saves any such theory from being immediately ruled out by the solar system observations,
  due to the fact that any such ghost-free theory must contain an extra scalar polarization  and 
  is subject to vDVZ discontinuity at the lineareazed level.  This fact makes impossible the existence
  of consistent  weakly coupled theories of modified gravity (relevant for cosmic observations).
  
  Interestingly, in the class of theories considered, the same scalar polarization that creates 
 the problem at the linearized level, also invalidates it by exhibiting the strong coupling phenomenon.  
 
 Due to this strong coupling phenomenon,  gravitating objects are `endowed' with a 
physical scale $r_*$, which for the extra graviton polarizations  plays the hole of a second   
`horizon'.  At distances $r_* \sim r$ the non-linear interactions of $\chi$ catch-up with the linear
part, and expansion in terms of the Newtonian coupling constant can no longer be trusted. Hence, 
vDVZ-discontinuity is invalidated.  

For scales $r \ll r_*$, the metric can be found in series of $\left ({r\over r_c}\right )^{2(1-\alpha)}$ - expansion (or possibly exactly). The assumption that the true solution can be approximated in such analytic  series, fixes the form of the leading correction to the Eisteinian metric in the form of 
(\ref{centralcor}).  

This fact, makes theories of modified gravity potentially testable by the precision gravitational experiments in the solar system, and in particular by Lunar Ranging experiments.  

Another important question is predicting the explicit form of modified Friedmann equation  for 
generic $\alpha$, in the spirit of \cite{dt,roman}.

$~~~~~~~$

$~~~~~~~$

{\bf Note Added}

  In this note we wish to briefly comment  on the issue of stability of some cosmological backgrounds, 
 that are created by large distance modification of gravity, since  this issue is governed  by the  
 concept of $r_*$, which is central to our work.  For the phenomenologically most interesting case $r_c \sim H_0^{-1}$, all the gravitating sources, that fit within the present Hubble volume, are smaller 
 than their own $r_*$-s.  The only observable source that is of the size of  its own $r_*$, is the Universe 
 itself.  This fact immediately explains why the smaller-than-$H_0^{-1}$ sources gravitate almost normally, whereas the cosmological expansion of the Universe is strongly modified. In \cite{dgp} the result 
 of the latter modification is the appearance of the self-accelerated background\cite{cd},\cite{ddg}. 
 Lately,  the perturbative stability of this background was subject of some discussion.
Our analysis indicates that the issue can only be answered non-perturbatively.  Indeed, since for the 
Universe $r_* \, \sim  \, H_0^{-1}$, the perturbation theory breaks down.  Another indication of this 
breakdown is that on the self-accelerated background  $\square \chi \sim  H_0^2/M_P$,
 which the reader  can easily  check.     
This simple fact is hard to reconcile with  the  claims  of \cite{ghost}, about the existence of ghost-type instabilities on this background, since these are based on the perturbative treatment, which  is invalid.     
The detailed discussion of the recent article\cite{cedricgiga}  points out yet another problem with the latter analysis.  The key point being, that for understanding the dynamics of  excitations,  that are localized  within their own $r_*$-s, again, the perturbative arguments cannot be trusted.

$~~~~~~$

  {\bf Acknowledgments}

We thank C.~Deffayet, G.~Gabadadze and M.~Redi for discussions. 
The work  is supported in part  by David and Lucile  Packard Foundation Fellowship for  Science and Engineering, and by NSF grant PHY-0245068. 
We also thank  Galileo Galilei Institute for Theoretical Physics and Institut des Hautes Etudes Scientifiques  for the hospitality and INFN for partial support during the completion of this work.


\end{document}